\documentclass[aip,jcp,reprint,graphicx]{revtex4-1}
 \usepackage{epsfig}
 \usepackage{amsfonts}
 \usepackage{color}
 \usepackage{amsmath}
\usepackage{graphicx}

\begin{document}

\title{Dynamic heterogeneity in an orientational glass}

\author{Nirvana B. Caballero }
\email{nirvana.caballero@cab.cnea.gov.ar  }
\affiliation{CNEA, CONICET, Centro Atomico Bariloche, 8400 San Carlos de Bariloche, Rio Negro, Argentina}

\author{Mariano Zuriaga }
\email{zuriaga@famaf.unc.edu.ar}
\affiliation{Facultad de Matem\'atica, Astronom\'{\i}a, F\'{\i}sica y Computaci\'on,
Universidad Nacional de C\'ordoba, C\'ordoba, Argentina and IFEG-CONICET,
Ciudad Universitaria,
X5016LAE C\'ordoba, Argentina}

\author{Josep Llu\'is Tamarit}
\email{josep.lluis.tamarit@upc.edu}
\affiliation{Grup de Caracteritzaci\'o de Materials, Departament de F\'isica, EEBE and Barcelona Research Center
in Multiscale Science and Engineering, Eduard Maristany, 10-14, 08019 Barcelona, Catalonia, Spain}

\author{Pablo Serra}
\email{serra@famaf.unc.edu.ar}
\affiliation{Facultad de Matem\'atica, Astronom\'{\i}a, F\'{\i}sica y Computaci\'on,
Universidad Nacional de C\'ordoba, C\'ordoba, Argentina and IFEG-CONICET,
Ciudad Universitaria,
X5016LAE C\'ordoba, Argentina}

\begin{abstract}

Family of compounds CBr$_n$Cl$_{4-n}$ has been proven helpful in unraveling
microscopic mechanisms responsible of glassy behavior. Some of the family members
show translational ordered phases with minimal disorder which appears to reveal glassy
features, thus deserving special attention in the search for universal glass anomalies.
In this work, we studied CBrCl$_3$ dynamics by performing extensive molecular
dynamics simulations. Molecules of this compound perform reorientational discrete jumps,
where the atoms exchange equivalent positions among each other revealing a cage-orientational jump motion fully comparable to the cage-rototranslational jump motion in supercooled liquids.
Correlation times were calculated from rotational autocorrelation
functions showing good agreement with previous reported dielectric results.
From mean waiting and persistence times calculated directly
from trajectory results, we are able to explain which microscopic
mechanisms lead to characteristic times associated
with $\alpha$ and $\beta$-relaxation times measured experimentally.
We found that two nonequivalent groups of molecules have a longer characteristic
time than the other two nonequivalent groups, both of them belonging to the
asymmetric unit of the monoclinic (C2/c) lattice.
\end{abstract}

\maketitle

\section{Introduction}

Glassy dynamics are present in all physical systems with non-stationary
dynamical processes in observable time scales, where due to competitive
interactions, the system remains trapped in metastable states.
Such dynamical processes are usually present in systems with a large
number of metastable configurations. At present, this behavior has been 
observed in quite diverse systems, like granular materials~\cite{josserand2000},
biologic evolution models~\cite{anderson2004evolution},
elastic strings in disordered media~\cite{bustingorry2007out},
domain wall dynamics~\cite{gorchon2014pinning}
and many substances and mixtures of substances of highly relevant technological
applications~\cite{angell1995formation}.

The glass and glass transition concepts were developed from the study
of supercooled liquids. However, some of their properties were found in
other condensed matter systems. In fact, detailed theoretical considerations
of molecular liquids~\cite{schilling2000mode} and some kind of molecular
crystals~\cite{ricker2005microscopic} have proven useful to establish
connections with theoretical results derived from several statistical
mechanics approaches where glassiness appears in the dynamics of idealized objects.
Such model systems show trivial, non-interacting, equilibrium behaviors but nonetheless, 
interesting slow dynamics features appear due to restrictions on the allowed
transition between configurations~\cite{Ritort_2003}.

Disordered molecular crystals known as rotator-phases or plastic-crystals
are simple systems (commonly with cubic lattice symmetry)
exhibiting dynamical behaviors which may at least partially
be understood by recourse to results obtained from idealized models~\cite{Vispa_PRL_2017}.
In these kind of systems, molecules may reorient by almost free rotations at the nodes
of a crystal lattice and by cooling such a rotator-phase,
a frozen orientationally disordered crystalline state
(called glassy crystal or orientational glass) may be 
attained, exhibiting fully developed glassy behavior.
Moreover, thermodynamic signatures similar to those common to 
vitrifying liquids are shown when approaching this state~\cite{Vispa_PRL_2017,Suga1974,Rute2004}.

There is also the possibility that a phase transition to a lower symmetry
crystal occurs, involving a partial reduction of the orientational disorder,
but keeping some occupational (low-dimensional) disorder.
These systems~\cite{Romanini_2012,Vdovichenko2015,Perez2015,BenHassine2016,Romanini2016,Tripathi2016,angell1996}
can still display glassy features and
represent the most simplified models on which theoretical concepts
on glassy dynamics can be tested.
Materials composed by tetrahedral molecules of general formula
CBr$_n$Cl$_{4-n}$, $n=0,\dotsc,4$ exhibit a series of solid-solid
phase transitions with increasing temperature before melting 
attributed to the ability of the molecules to acquire rotational
degrees of freedom as the temperature is increased~\cite{Pardo_PCCP_2001,Parat_2005,Barrio2008}.
Cooling the room temperature liquids leads to rotationally disordered crystalline
phases (rotator or plastic phase) which show translational face-centered
cubic or rhombohedral lattices with the carbon atoms sitting
at the lattice nodes. Further cooling leads to a transformation into complex
monoclinic, C2/c structures with Z=32 molecules per unit cell and an asymmetric
unit with Z'=4~\cite{Pardo_JCP_1999,Pardo2000stable,Pardo_PCCP_2001,Pardo_JPCB_2001,Parat_2005,Barrio2008,Pothoczki_2012}.

Molecular motions within such
monoclinic crystals persist down to $\sim$90~K where a calorimetric transition much
alike that exhibited by the canonical glass-transition signals the transition into
a orientationally disordered state where molecular reorientations cannot be detected
with the available experimental means~\cite{Ohta_1995}.
The dynamics of the monoclinic phases of CBr$_n$Cl$_{4-n}$, n=0,1,2 compounds 
has been studied by means of dielectric spectroscopy and nuclear quadrupole
resonance (NQR) spectroscopy~\cite{Zuriaga_PRL_2009,Zuriaga_JCP_2012} in the
temperature range 100 -- 250~K and 80 -- 210~K, respectively.
The former technique allows the measurement of the dynamic response within a broad time
scale but it is insensitive to fine details of molecular motions, whereas the latter has
a restricted time window but monitors the movement of individual chlorine atoms.
Results derived from the concurrent use of NQR and molecular simulations show that
large-angle rotations of tetrahedral about their higher symmetry axes (CBrCl$_3$,
and CBr$_2$Cl$_2$ with $C_{3v}$ and $C_{2v}$ point-group symmetries, respectively)
lead to a statistical occupancy of 75$\%$ for Cl and 25$\%$ for Br atoms in
the case of CBrCl$_3$, and 50$\%$ for Cl and 50$\%$ for Br atoms for CBr$_2$Cl$_2$
in agreement with X-ray and neutron diffraction measurements~\cite{Lee_1997,Binbrek_1999,Parat_2005,Barrio2008}.
These experimental techniques also show that the relaxation movements arise
from different dynamics exhibited by molecules in the
asymmetric unit of the crystalline lattice that are non-equivalent with respect to
their molecular environment.

The dielectric spectra of CBrCl$_{3}$ and
CBr$_{2}$Cl$_{2}$ at the lower end of the temperature range
display a well-defined shoulder on the high frequencies flank of the 
$\alpha$-peak, which is attributed to the $\beta$-relaxation~\cite{Zuriaga_PRL_2009}.
Since CCl$_4$ has no molecular dipole moment it is not
accessible to dielectric experiments but it can be studied using NQR.
Interestingly, the resolution of the
NQR spectra for CCl$_4$ is well superior to the
corresponding spectra for CBr$_{2}$Cl$_{2}$ and
CBrCl$_{3}$ and the two techniques complement each other.
On the other hand, the NQR experiments are limited to a temperature
range between 77~K and 140~K, with the upper end determined by the
broadening of the signal~\cite{Zuriaga_JCP_2012}.
The picture that emerges from the combined
analysis is that the three compounds have a very similar dynamic
evolution in the monoclinic phase as a function of temperature
\cite{Zuriaga_PRL_2009,Zuriaga_JCP_2012}.
The analysis of the isostructural CCl$_4$ shows that nonequivalent
molecules in the unit cell perform reorientational
jumps at different time scales due to their different
crystalline environments.
These results support the conclusion that the dynamic heterogeneity
is intimately related to the secondary relaxation observed
in these compounds~\cite{Zuriaga_PRL_2009,Zuriaga_JCP_2011,Zuriaga_JCP_2012}.

The currently accepted scenario for canonical glasses includes
different relaxation mechanisms that are universally present in all
systems. Experimentally, these different mechanisms are enclosed into
the dielectric spectra that show a broad low frequency peak
referred to as $\alpha$-relaxation~\cite{Lunkenheimer_2002} and a higher frequency peak
or shoulder usually called Johari-Goldstein $\beta$-relaxation~\cite{Johari_1970,Johari_1973,Ngai_1998,JimenezRuiz_PRL_1999,JimenezRuiz_PRB_1999,Schneider_2000,Johari_2002,Ngai_2004,Affouard_2005,Pardo_2006,Capaccioli_2007,Capaccioli_2011}.
The $\alpha$-relaxation is due to processes involving
cooperative dynamics of regions of molecules~\cite{Lunkenheimer_2002}.
The microscopic origin of the $\beta$ relaxation is still a matter of wide debate~\cite{Johari_1970,Johari_1973,Ngai_1998,JimenezRuiz_PRL_1999,Schneider_2000,Johari_2002,Ngai_2004,Pardo_2006,Capaccioli_2007,Capaccioli_2011,Roland2017}.
Some of the proposed models explain this peak as a consequence of the non-uniformity
of the glassy state involving only local regions in
which molecules can diffuse (islands of mobility).
An alternative homogeneous explanation attributes the
secondary relaxation phenomena to small-angle reorientations of all
molecules~\cite{Romanini_2012,Voguel_2000,Voguel_2001}.

In our previous work on CCl$_4$~\cite{Caballero_2016} we found
that the monoclinic phase of this compound has
essentially the same dynamical
behavior, as function of temperature, as that of its
isostructural glass formers, CBrCl$_3$ and CBr$_2$Cl$_2$.
The simulations clearly show
that there are preferential axes of rotation, which are
fixed with respect to the crystal orientation.
Two of the inequivalent groups of molecules among
the four non-equivalent molecules of the asymmetric
unit are significantly faster than the other two,
leading to a clear heterogeneity in the
dynamics of the system. Moreover, it is found that the orientation
of the two fast axes of rotation is the same, suggesting an
overall dynamics anisotropy correlated to the molecular
orientations. We showed that the different reorientational dynamics
in CCl$_4$ (not involving structural changes), are responsible of characteristic
times compatible with $\alpha$ and $\beta$-relaxations times, measured
with dielectric spectroscopy for the members of the family which can
orientationally vitrify.

In this work we have studied the monoclinic phase of
CBrCl$_3$ using molecular dynamics simulations in order
to elucidate the mechanisms responsible of the $\alpha$ and $\beta$-relaxations
observed experimentally \cite{Zuriaga_PRL_2009,Zuriaga_JCP_2012}.

\section{Theoretical Methods}
\subsection*{Model and computational details}

CBrCl$_3$ is a non-regular tetrahedral molecule with $C_{3v}$ molecular
symmetry. We have
modeled CBrCl$_3$ molecules as rigid, non-polarizable tetrahedra with
the carbon atom at the center, three chlorine atoms on three vertices and
a bromine atom located in the remaining vertex, as was proposed
in our previous work\cite{Caballero2012}.
The interaction between
molecules is represented by a combination of Lennard-Jones and Coulombic
terms summarized in Table \ref{table:potential}. The cross interaction
between atoms of different type is calculated by applying the
Lorentz-Berthelot combination rules, i.e., geometrical mean for $\epsilon$
and arithmetic mean for $\sigma$. A spherical cut-off at 1.8 nm was imposed
on all intermolecular interactions. Periodic boundary conditions were imposed
in all three Cartesian directions.

\begin{table}[ht]
\caption{\label{table:potential} CBrCl$_3$ model parameters and geometry.}                        
\centering  
\begin{tabular}{l c c c c c}
\hline\hline \\[-2.0ex]                                                                                                                                      
  & $\epsilon$ [kJ/mol] & $\sigma$ [nm] & $q$ [e] & \multicolumn{2}{c}{Bond [nm]} \\ 
\hline
C & 0.22761 & 0.37739 & -0.696 & C-Cl & 0.1766 \\
Cl& 1.09453 & 0.34667 & 0.174  & Cl-Cl& 0.2884 \\
Br& 2.13000 & 0.37200 & 0.177  & C-Br & 0.1944 \\
\hline\hline
\end{tabular}
\end{table}

The isostructural series CBr$_n$Cl$_{4-n}$ have a low temperature monoclinic
crystal structure, resolved by Cohen et {\em al.} at 195~K~\cite{Cohen1979}
which corresponds to the $C2/c$ space group. The unit cell, with $Z$=32
molecules, has the following lattice parameters:
$a$=2.0631~nm, $b$=1.1619~nm, $c$=2.0201~nm and
angle $\beta=111.19^{\circ}$ at 220.2~K~\cite{Negrier_2007}.
Using the experimental crystalline structure as initial coordinates,
we constructed a simulation super-cell containing  512 molecules,
which correspond to 16 monoclinic unit cells. This super-cell was prepared by replicating
the experimental unit cell twice on the $x$ and $z$ directions and four times in the $y$
direction.

The molecular dynamics simulations, conducted under NPT conditions, have been carried out
using the Gromacs v5.0.2 simulation package. Atom-atom distances within each molecule
were kept constants with the SHAKE algorithm. The classical Newton's equations were
integrated using the leap-frog algorithm and the time step of the integration of the
equations of motion was set to 5~fs. 

The production runs were extended up to 200~ns or 11~$\mu$s depending on the temperature.
The temperature control was implemented with a Nos\'e-Hoover thermostat, with a time
constant of 2.0~ps. The pressure was maintained constant by using a fully anisotropic
Parrinello-Rahman barostat with a reference pressure of 1~atm. The study covered
temperatures ranging from 160~K to 220~K, in steps of 10~K.

\section{Results}

Molecular dynamics simulations of CBrCl$_3$ for the whole range of studied temperatures capture the essential dynamics of CBrCl$_3$ proposed to the moment: we observe a monoclinic structure, where carbon atoms are centered in the lattice nodes and all molecules perform reorientational jumps between equilibrium positions.

Molecular rotations were characterized by angular self-correlation functions defined as~\cite{kammerer1997dynamics} 

\begin{equation}
C^{b}(t)=\frac{1}{N}\sum_i^N \langle \vec{u}_i^{b}(\zeta)\cdot\vec{u}_i^{b}(t+\zeta) \rangle_\zeta, 
\label{eq:correlations}
\end{equation}

\begin{figure}[ht]
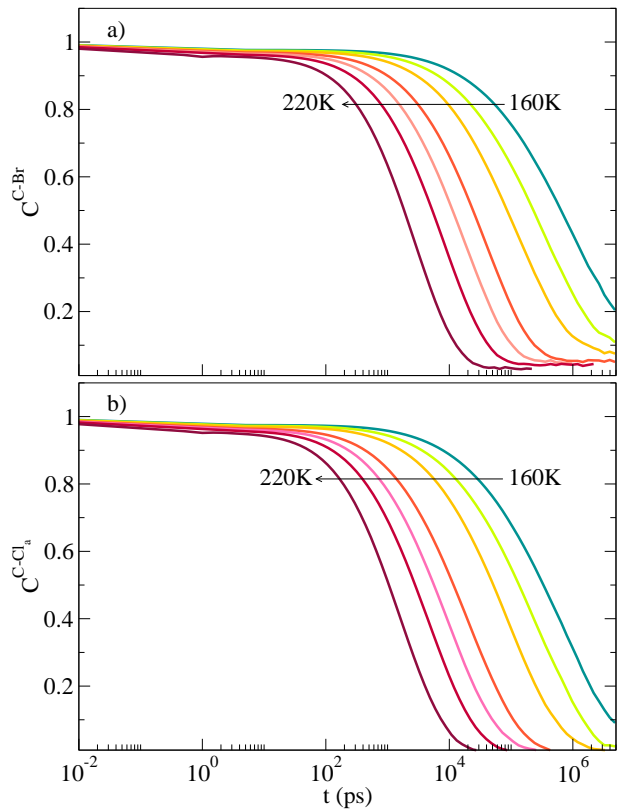

\begin{center}
\centering
\includegraphics[width=0.45\textwidth]{rot_cbr_T_prom.eps}
\includegraphics[width=0.45\textwidth]{rot_ccla_T_prom.eps}
\end{center}
\caption{Rotational correlation functions for all a) C-Br and b) C-Cl$_a$ directions for temperatures ranging from 160~K to 220~K in steps of 10~K.}
\label{fig:rot}
\end{figure}

\begin{figure}[ht]
\begin{center}
\centering
\includegraphics[width=0.45\textwidth]{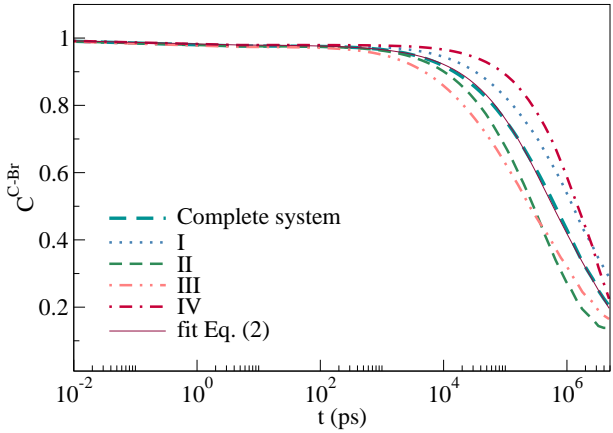}
\end{center}
\caption{Rotational correlation functions for C-Br bonds at 160~K, calculated for molecules belonging to the four different groups in the system and for the whole system. The fit of the rotational correlation function for the complete system according to Eq.~(\ref{eq:fitcorrelations}), with $C_1=0.68$, $\beta_1=0.54$, $C_2=0.3$, $\beta_2=0.72$, $\tau_1=3.2\times 10^6$~ps and $\tau_2=2.4\times 10^5$~ps is also included. Correlation functions of groups I and IV are similar and the same holds for groups II and III (this behavior was observed for all studied temperatures). This coincidence among correlation functions is the first signature of two main dynamics: the faster one, where molecules of groups II and III are involved and the slower one, involving molecules of groups I and IV.}
\label{fig:rotgroups}
\end{figure}

\noindent where, $\vec{u}_i^{b}$ is the normalized vector directed along one of the CBrCl$_3$ intra-molecular bonds $b$=C-Br, C-Cl$_a$, C-Cl$_b$, C-Cl$_c$, $i$ is the molecule number, and $N$ is the number of molecules considered in the calculation. The average is over times $\zeta$. These self-correlation functions were calculated for the four molecular bonds for all molecules in the system. We found the same behavior for all three C-Cl bonds, but a slower dynamic for the C-Br bonds. The functions are shown for bonds $b$=C-Br and $b$=C-Cl$_a$ in figure \ref{fig:rot} for the whole range of studied temperatures.

CBrCl$_3$ is isostructural to CCl$_4$. The CCl$_4$ unit cell is monoclinic,
with Z=32 molecules per unit cell and space group C2/c.
The spatial positions of the molecules in the unit cell
are defined through the application of 8 symmetry operations over Z'=4
nonequivalent molecules. As a consequence, the system has four distinctive
groups of molecules (according to the Z'=4 molecules in the asymmetric
unit and Z=32 molecules in the unit cell) that we will refer to as
groups I, II, III and IV. In our simulation cell, each group contains 128 molecules.
In the CCl$_4$ system, each molecule belonging to a given
group has the same specific arrangement of neighboring molecules.
However, in a CBrCl$_3$ system, the bromine atoms
introduce disorder in the system and molecules belonging to the same group
have no longer the same environment.

In order to get deeper understanding of the system rotational dynamics,
we have also calculated the self-correlation functions of molecules
belonging to the four non-equivalent groups.
These functions are shown in figure~\ref{fig:rotgroups} for T=160~K.
We found that groups I and IV have similar dynamics, and the same
conclusion holds for groups II and III, for all studied temperatures.
Here, molecules in groups II and III have faster reorientational
dynamics than molecules belonging to groups I and IV.
The two different dynamics of nonequivalent molecules
in the asymmetric unit was previously detected by means
of dielectric and NQR techniques~\cite{Zuriaga_PRL_2009, Zuriaga_JCP_2012}
and was also observed by us for CCl$_4$~\cite{Zuriaga_JCP_2011,Caballero_2016},
where the same groups of molecules were classified as slow (I and IV)
and fast (II and III).

We have calculated relaxation times associated with all bromine atoms
in the system from self-correlation functions (Eq.~\ref{eq:correlations}),
since them can be directly compared with experimental results based
on dielectric techniques, reported earlier~\cite{Zuriaga_PRL_2009} 
(because the C-Br bond defines the molecular dipole direction).
The contribution to the relaxation spectrum from $\alpha$ and $\beta$-processes
may be deconvoluted using the William ansatz~\cite{Williams,Roland2014,Roland2017}

\begin{equation}
C(t)=C_1 e^{(\frac{-t}{\tau_1})^{\beta_1}} + C_2 e^{(\frac{-t}{\tau_2})^{\beta_2}}.
\label{eq:fitcorrelations}
\end{equation}

\begin{figure}[ht]
\centering
\includegraphics[width=0.45\textwidth]{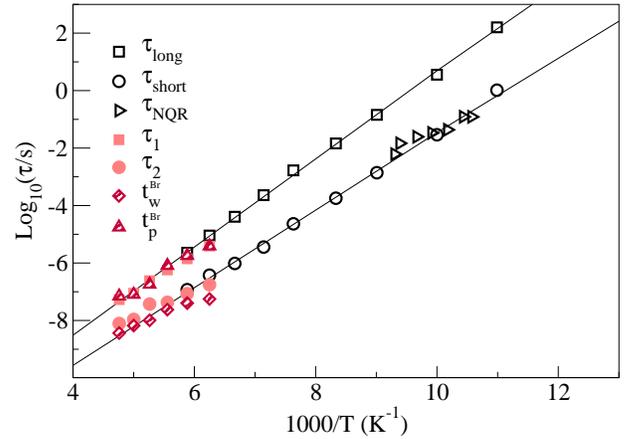}
\caption{\label{fig:taus} Spin-lattice relaxation times obtained by
molecular dynamics simulations from C-Br self-correlation functions (full squares and circles).
Waiting (diamonds) and persistence (triangles) mean times are also shown.
$\tau_{long,short}$ (empty squares and circles) are relaxation times obtained
by dielectric means for CBrCl$_3$, taken from Ref.~\cite{Zuriaga_PRL_2009}.
Black straight lines are fits of $\tau_{long}$ and $\tau_{short}$.
Results from NQR studies of CBrCl$_3$~\cite{Zuriaga_PRL_2009}
(empty triangles) are also included.}
\end{figure}

The self-correlation functions for all C-Br bonds in the system are
well fitted with a combination of two stretched exponentials.
The Kohlrausch-Williams-Watts (KWW),
or stretched exponential decay function is the most commonly used empirical decay
function for handling relaxation data affected by disorder~\cite{Richert_2012}.
It is very successful in describing relaxation data of many disordered systems,
since it is capable of capture the large range of rate constants involved.
We found, for the whole range of studied temperatures, $0.63 \le C_1 \le 0.74$,
$0.54 \le \beta_1 \le 0.71$ (associated with the longer relaxation time $\tau_1$) and
$0.24 \le C_2 \le 0.35$, $0.72 \le \beta_2 \le 0.8$
(associated with the shorter relaxation time $\tau_2$).
Relaxation times $\tau_1$ and $\tau_2$, associated with all the bonds C-Br
in the system, showing a very good agreement with
experimental results, are plotted in figure~\ref{fig:taus}
along with dielectric and NQR results taken from Ref.~\cite{Zuriaga_PRL_2009}.
$\tau_1$ is compatible with $\tau_{long}$ and $\tau_2$ with $\tau_{short}$.
Dielectric spectra for CBrCl$_3$, as was reported earlier~\cite{Zuriaga_JCP_2012},
is very well described by a Havriliak-Negami function for the $\alpha$-processes.
From this function, a $\beta_{D}$, equivalent to the stretching exponent in a decay 
behavior, is obtained. The stretching parameters $\beta_2$ obtained by fitting the self-correlation functions of C-Br bonds are similar to $\beta_D$.

\begin{figure}[ht]
\begin{center}
\includegraphics[width=0.45\textwidth]{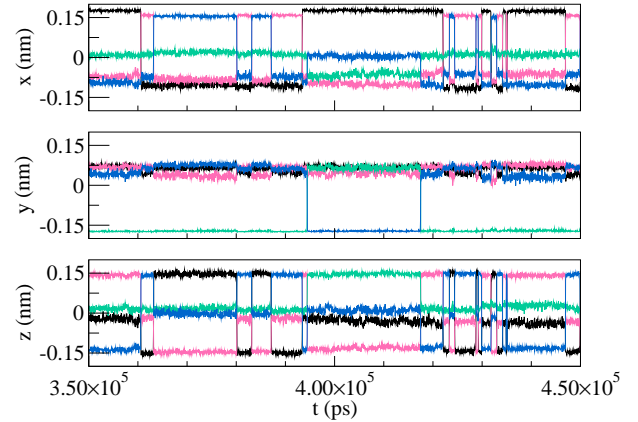}
\end{center}
\caption{\label{fig:coordinates} 
Cl and Br coordinates relative to the C atom for one arbitrary molecule of group II at 160~K. Black, pink, green and blue, are for Br, Cl$_a$, Cl$_b$ and Cl$_c$ respectively.}
\end{figure}

The molecular reorientation processes occurring during the simulations
are sudden large-angle jumps of Cl and Br atoms.
In figure~\ref{fig:coordinates}, coordinates of Br and Cl atoms of
one molecule arbitrarily chosen are shown as function of time during 100~ns.
The C atom is in a lattice node, and the remaining atoms of the molecule
exchange equilibrium positions with each other: two types of jumps are
possible in this situation, one in which all atoms exchange their
previous average position, i.e four atoms jump (C$_2$ jumps),
and other where an atom keeps its average position,
and the remaining three exchange positions between them (C$_3$ jumps).

It is known that in the case of molecular liquids and hard-sphere-like
colloidal glasses, on approaching the
glass transition to an amorphous solid-like state, the dynamic becomes
intermittent and shows large spatiotemporal fluctuations, also known as dynamic
heterogeneities~\cite{Pastore2014,Pastore2017}. The dynamic is spatiotemporal heterogeneous
since the probability that a particle rearranges in a given time interval
is not spatially uniform, as long as the considered time interval is smaller
than the relaxation time. Glassy dynamic is intermittent, as particles suddenly
jump out of the cage formed by their neighbors, and heterogeneous,
as these jumps are not uniformly distributed across
the system~\cite{Pastore2015,Pastore2017}. 
Since particle jumps are an ubiquitous feature of supercooled
liquids~\cite{Ciamarra2016}, it is interesting to consider their role in an
glass system in which only orientational dynamics can be frozen, as CBrCl$_3$.

\begin{figure}[ht]
\begin{center}
\includegraphics[width=0.45\textwidth]{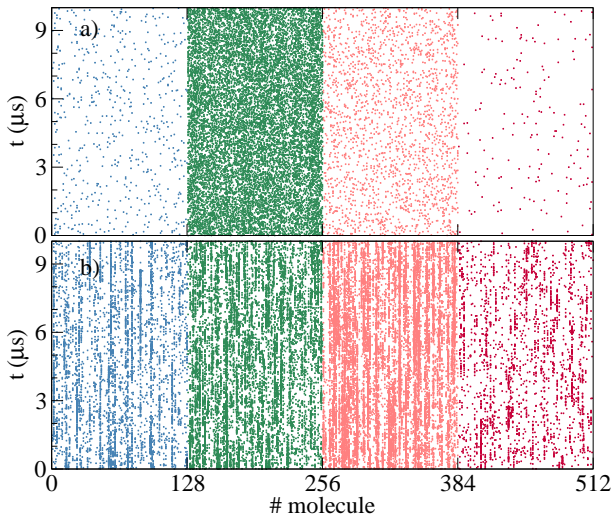}
\end{center}
\caption{\label{fig:jumps} 
Registered jumps for a) CCl$_4$ and b) CBrCl$_3$ for every molecule of
the system during a simulation of 10~$\mu s$ at 160~K.
Molecules numbered from 1 to 128, 129 to 256, 257 to 384
and 381 to 512 belong to groups I, II, III and IV respectively.
First 8 molecules of every group are in the same unit cell.
Second 8 molecules are in another unit cell and so on.
CCl$_4$ shows a Poisson distribution of jumps as was highlighted
before~\cite{Caballero_2016}, but CBrCl$_3$ moves away from this
behavior. For the CCl$_4$ system, a clear difference in number
of jumps among groups arise, while in the CBrCl$_3$ system, there
is not such a clear distinction. In the CBrCl$_3$ system is worth
to notice that molecules with short waiting time between jumps are
more probable to have a next short waiting time.}
\end{figure}

The jumps for all molecules were detected using a running test
algorithm based on a signal to noise measure described in our
previous work for CCl$_4$~\cite{Caballero_2016}.
After analyzing the trajectories for the $N$=512 molecules in the system,
the times $t^{n,i}$ at which every single jump occurs were registered. 
The index $i$ indicates the $i^{th}$ jump of the $n^{th}$-molecule,
$n=1,\dotsc,N$. The method of jump detection also allows to distinguish
which atoms of each molecule are involved in the jump process.
A careful inspection confirmed that for all temperatures
there were just a handful of cases corresponding to $C_2$
type rotations and therefore there were neglected in the analysis.

In figure~\ref{fig:jumps} all registered times $t^{n,i}$
for CBrCl$_3$ at 160~K are shown.
The molecule number $n$ is such that molecules
$n=1,\dotsc,128$, $n=129,\dotsc,256$, $n=257,\dotsc,384$
and $n=385,\dotsc,512$ belong to groups I, II, III and IV, respectively.
In order to compare with CCl$_4$, the times at which
jumps occur for this compound are also plotted at the
same temperature, with data taken from our previous
work~\cite{Caballero_2016}.

A feature of figure~\ref{fig:jumps} is important to highlight:
the distribution of jumps for CCl$_4$ is uniform in time,
while for CBrCl$_3$ some correlations arise.
Periods of fast and slow reorientations are noticeable,
indicative of heterogeneous dynamics.

\begin{figure}[ht]
\begin{center}
\includegraphics[width=0.5\textwidth]{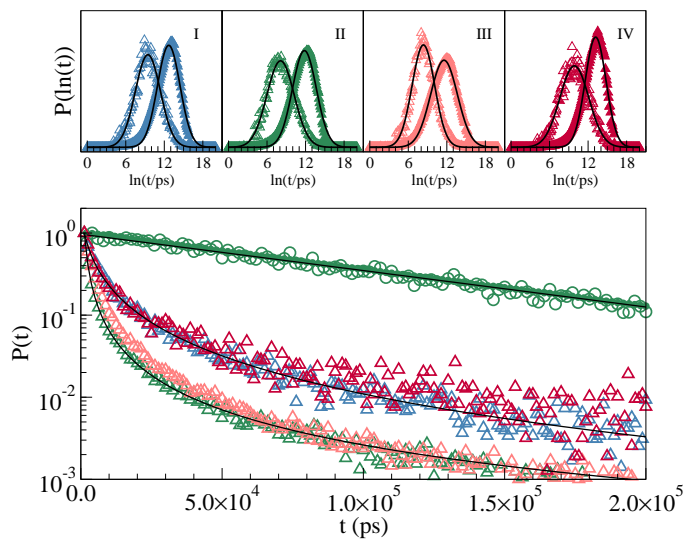}
\end{center}
\caption{\label{fig:wt}
Probability distributions of waiting times (empty triangles)
for groups I (blue), II (green), III (pink) and IV (red)
of CBrCl$_3$. Probability distributions of waiting times
(empty circles) and  persistence times (full circles)
for group II of CCl$_4$ are also included.
In the top figures the probability distributions of the natural
logarithm of waiting times (empty triangles) and persistence times
(full triangles) for CBrCl$_3$ are plotted.
All data is at 160~K. Black solid lines are curves resulting
from fitting the data with a normal function for the four top figures,
an exponential function for CCl$_4$, and a log-normal function for
CBrCl$_3$ (just fits of groups I and II are shown for CBrCl$_3$).}
\end{figure}

The statistics behind the rotational jumps can be evaluated
by studying the waiting times between successive jumps.
We define the waiting time $t^{n,i}_w$ as the time between
two consecutive jumps of the same molecule,
i.e. $t^{n,i}_w$=$t^{n,i+1}-t^{n,i}$.
In figure \ref{fig:wt} we show the distribution of waiting times
for rotational events for CBrCl$_3$ and CCl$_4$ directly obtained
from the simulated trajectories.
For CCl$_4$ the waiting times distribution is a Poisson process,
following an exponential law $\sim e^{t/\tau_{w}}$.
Calculating the probability distribution of $x=ln(t_w)$,
a normal distribution, $\sim e^{\frac{(x-\mu)^2}{2\sigma^2}}$
results suitable to fit the data in the case of CBrCl$_3$.
A standard variable change procedure gives a log-normal distribution
for the probabilities as function of time, with mean equal
to $\tau_w=e^{\mu+\frac{\sigma^2}{2}}$.

\begin{figure}[t]
\begin{center}
\includegraphics[width=0.5\textwidth]{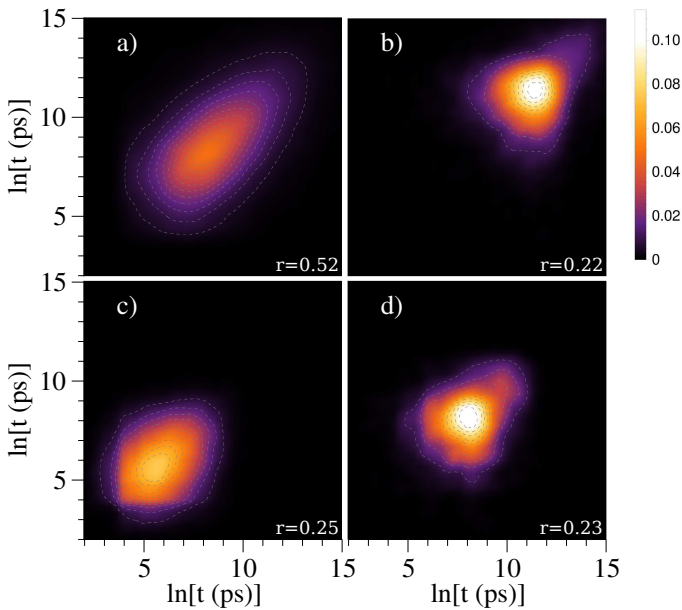}
\end{center}
\caption{Probability distribution functions and contour plots (dashed lines)
showing the dependence between subsequent waiting times for all molecules in a system of a) CBrCl$_3$ molecules at 160~K, b) $CCl_4$ molecules at 160~K, c) CBrCl$_3$  at 220~K and d) CCl$_4$  at 220~K. Vertical axis and horizontal axis indicate waiting times and previous waiting time, respectively.}
\label{fig:correlationswt}
\end{figure}

The dependence of one waiting time ($t^{n,i}_w$) on the previous
waiting time ($t^{n,i-1}_w$) provides a measure for the heterogeneities
lifetime~\cite{Bingemann2011}.
In the case of homogeneous dynamics for one molecule,
subsequent waiting times are uncorrelated.
For very long heterogeneity lifetimes,
a long waiting time is most likely to be followed
by another long waiting time, while a short waiting time
is most likely followed by another short waiting time.
In order to analyze correlations between subsequent
waiting times we have calculated the contour plots shown in
figure~\ref{fig:correlationswt}.
These contour plots are constructed from the probability
distribution of having a $t^{n,i}_w$ considering that
the previous waiting time was $t^{n,i-1}_w$.
The analysis was made for the CBrCl$_3$ and CCl$_4$ systems and is shown
in figure~\ref{fig:correlationswt}, at the highest and lowest simulated temperatures.
In the CBrCl$_3$ case, at 160~K, an elongated cloud along the diagonal
leads to a correlation of $r$=0.52 between the logarithm of subsequent
waiting times; while for the highest simulated temperature,
220~K, this value is reduced to 0.25, almost
half of the previous case, closer to the CCl$_4$ case, where correlations
are 0.22 and 0.23 at 160~K and 220~K, respectively.

One consequence of correlations among waiting times is that
dynamical processes can be partitioned according to whether they
coincide with waiting events (successive jumps of the same molecule)
or persistence events (the time from the molecule initial condition before
it jumps). The distinctions between these two classes of processes
are responsible for a host of nonlinear phenomena that are characteristic of deeply
supercooled liquids \cite{Chandler2010}.

The distribution of times characterizing the molecular intermittent
behavior can be viewed in terms of the persistence time, 
defined as the time from any given point in the trajectory 
of a single molecule $t^{i}_0$, until the next jump event $t^{n,j}$:
$t^{n,i}_p=t^{n,j}-t^{i}_0$ , and the
time between two jumps (the, defined before, waiting time).
For uncorrelated sequence of events, persistence times and waiting times
have the same distribution, and the probability that an event has not occurred
in a time $t$ is $e^{-t/\tau}$,
with $\tau=t_w=t_p$, were $t_w$ and $t_p$ are mean waiting
and persistence times, respectively.
Such intermittencies are known as Poisson processes.
In the case of glassy dynamics, however, statistics of events
is far from Poissonean because an event
becomes more likely when a similar event has occurred more recently.
As a consequence, when dynamics are correlated, waiting times are 
typically shorter than persistence times \cite{Chandler2010,Hedges2007}.

\begin{figure}[h]
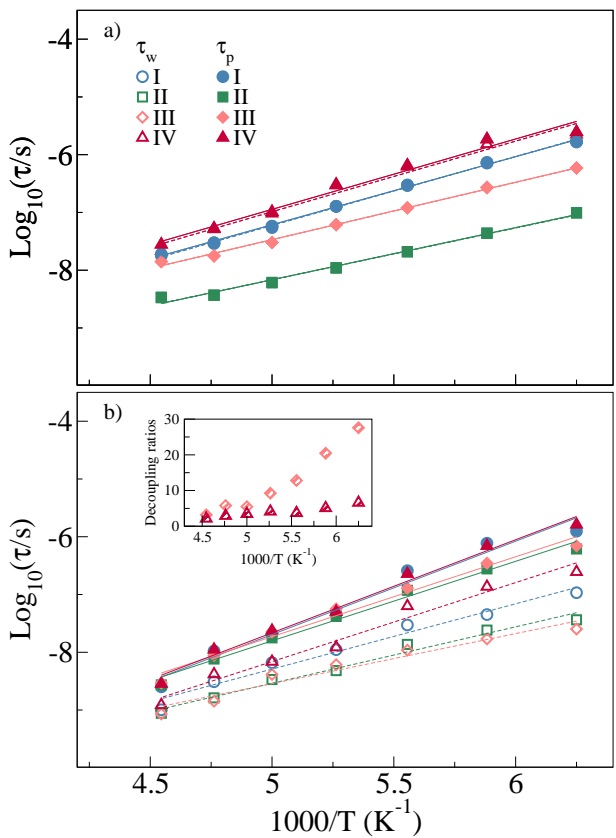

\begin{center}
\includegraphics[width=0.45\textwidth]{taus_ccl4.eps}
\includegraphics[width=0.45\textwidth]{taus_cbrcl3.eps}
\end{center}
\caption{\label{fig:tauswp}
Mean waiting (empty symbols) and persistence (full symbols) times obtained from probability distributions as function of temperature for the four groups (I, II, III and IV) in the system. Results for CCl$_4$ in a) show coincidence among mean waiting and persistence times for molecules in the same group. In the case b) for CBrCl$_3$, persistence mean times are longer than waiting mean times. In the inset of b) decoupling ratios, calculated as $t_p/t_w$ are shown for groups III (diamonds) and IV (triangles).}
\end{figure}

We have calculated persistence times distributions for CBrCl$_3$
and CCl$_4$, for all simulated temperatures and for the
four groups of nonequivalent molecules.
We found that for CBrCl$_3$ persistence
and waiting times distributions decouple (see figure~\ref{fig:wt}).
For CCl$_4$, the situation is different,
persistence and waiting times are equally distributed
and Poissonean. Mean values $\tau_w$ and $\tau_p$ for each group in
CBrCl$_3$ and CCl$_4$ are shown in figure~\ref{fig:tauswp}.
In the CCl$_4$ case, and impressive coincidence
among waiting and persistence mean times arises,
emphasizing the Poissonean behavior of the system.
For CBrCl$_3$, the correlation among jumps processes
is responsible of the marked difference between waiting 
and persistence mean times, where mean waiting times
are considerably shorter than mean persistence times.
In figure \ref{fig:tauswp} we also show the decoupling ratios, defined
as $\tau_p/\tau_w$, for groups III and IV. For the higher studied temperatures,
the ratio is close to one, as expected for standard liquid
conditions \cite{Chandler2010}, but for lower temperatures the ratio grows,
as in the fluctuation dominated regime of supercooled liquids, where this is
a signature of the breakdown of Stokes-Einstein relations \cite{Jung2004,Hedges2007}.

Persistence times, defined as the average waiting time
from any given point in the trajectory of a single molecule, 
until the next jump event, are comparable to the {\it structural}
relaxation time $t_{\alpha}$ in simple models of
glass dynamics \cite{ahn2013}. The probability
distribution of persistence times is comparable
to typical results of bulk experiments, due to the time-weighted
nature of typical observations.
Since available experimental times describing CBrCl$_3$ dynamics
are the result from dielectric measurements, where the bromine atom is
sensed, we also calculated persistence and waiting mean times for
all bromine atoms in the simulated system. 

We define the waiting
time for a bromine atom as the difference in time between successive
jumps of a bromine atom belonging to one molecule of the system.
The probability distributions for waiting times of all bromine atoms
in the system, as was the case for all molecular jumps, also
follow a log-normal distribution for all simulated temperatures.
The same behavior was observed for squared waiting times.
From mean values of these distributions, we calculated $\tau_w^{Br}$
and $\tau_p^{Br}$,
mean waiting and persistence times
for all bromine atoms in the system.
These results are plotted in figure \ref{fig:taus}.  
Persistence mean times of bromine atoms show an impressive good agreement with times $\tau_{long}$  obtained from dielectric measurements, and waiting mean times coincide with $\tau_{short}$ obtained in the same
experiment. The interpretation of this coincidence is that $\alpha$-relaxation
is associated with times in which bromine atoms are not reorienting, but persisting in one position,
and $\beta$-relaxation is associated with the fastest process in which the molecules are jumping and reorienting.

\section{Conclusions}

We performed extensive molecular dynamics simulations of CBrCl$_3$ in
the monoclinic low-temperature phase. By studying time reorientational auto-correlation
functions as function of temperature, we found distinct behaviors
for C-Br bonds when comparing with the other C-Cl bonds in the molecule.
We found two characteristic times emerging from these functions: a 
longer one, compatible with $\tau_{long}$ associated with $\alpha$-relaxation,
and a shorter one, compatible with $\tau_{short}$ associated with $\beta$-relaxation
from dielectric measurements.

CBrCl$_3$ dynamics reorientations are due to sudden jumps between
equivalent positions of the atoms. By computing jumping times
directly from simulation trajectories, we calculated waiting and
persistence times. Probability distributions of waiting and persistence times show
a log-normal behavior, from which is possible to obtain mean values for 
waiting and persistence times.

We calculated mean waiting and persistence times, first distinguishing
among groups of molecules, and second for all bromine atoms in the system.
From the groups analysis, we found that different groups have different
characteristic mean times, and when comparing mean waiting and persistence
times a difference arises, showing correlations in the system dynamics.
From the bromine atoms analysis, we found that mean waiting times are compatible
with the shorter characteristic time, associated with $\beta$-relaxation and
mean persistence times are compatible with the larger characteristic time,
associated with $\alpha$-relaxation.

Although groups of different molecules are
well defined for CCl$_4$, where every molecule belonging to
a group has the same molecular environment, in CBrCl$_3$,
where this assertion is not longer valid due to the disorder introduced
by the bromine atom, groups of molecules are still well defined for this compound.
Molecules belonging to different groups of the CBrCl$_3$ system
experiment different dynamics, as can be observed when studying autocorrelation
functions for bonds of molecules belonging to different groups,
or when the mean persistence and waiting times are calculated.

To our knowledge, this is the first time that an analysis in terms of
microscopic molecular jumps is done over an orientational low-dimensional glass. 
We showed that the orientational glass former CBrCl$_3$
has all the major features of structural glass formers, without of course,
the structural degree of freedom.

\section{Acknowledgements}

N.B.C., M.Z. and P.S. acknowledge financial support of SECYTUNC and CONICET.

This work used computational resources from
CCAD Universidad Nacional de C\'ordoba
(http://ccad.unc.edu.ar/),
in particular the Mendieta Cluster, which is part
of SNCAD MinCyT, Rep\'ublica Argentina.

This work was partially supported by the Spanish Ministry of Economy and Competitiveness through the project FIS2014-54734-P
 
\bibliography{biblio}

\end{document}